# Multi-linear iterative $K$-$\Sigma$-semialgebras

Z. Ésik[*]

Dept. of Computer Science
University of Szeged
Árpád tér 2
6720 Szeged, Hungary

**Abstract**

We consider $K$-semialgebras for a commutative semiring $K$ that are at the same time $\Sigma$-algebras and satisfy certain linearity conditions. When each finite system of guarded polynomial fixed point equations has a unique solution over such an algebra, then we call it an iterative multi-linear $K$-$\Sigma$-semialgebra. Examples of such algebras include the algebras of $\Sigma$-tree series over an alphabet $A$ with coefficients in $K$, and the algebra of all rational tree series. We show that for many commutative semirings $K$, the rational $\Sigma$-tree series over $A$ with coefficients in $K$ form the free multi-linear iterative $K$-$\Sigma$-semialgebra on $A$.

## 1 Introduction

In an iterative algebra [1, 9, 22, 19], certain finite systems of fixed-point equations have unique solutions. It was first pointed out in [19] that iterative algebras form a quasi-variety, hence all free iterative algebras exist. Clearly, the same holds for the subclass of iterative algebras in any variety or quasi-variety of algebras. However, it is usually a nontrivial task to find a concrete description of the free algebras in a certain class of iterative algebras.

Free algebras in the class of all iterative algebras may be represented as algebras of regular (finite or infinite) trees, cf. [22]. A characterization of the free iterative idempotent semirings by regular word languages is implicit in Salomaa's axiomatization of regular languages [20]. Extensions of Salomaa's axiomatization to rational power series (or weighted rational languages) are given for fields in [18], for commutative rings in [15], and for commutative "proper semirings" (including all finite commutative semirings and commutative rings) in [11]. It follows from these results that in certain classes of iterative semirings or iterative semialgebras, the free algebras may be represented as algebras of rational power series. (Some results along this line are claimed in the papers [4, 23], but the proofs are incomplete, as pointed out in the MR review of the second paper.) As an important consequence of the concrete representation, one can derive several decidability and undecidability properties of the corresponding equational theories and use automata theoretic arguments to establish the validity of equations. For example, it follows that the equational theory of the iterative algebras of [22, 19] is decidable in polynomial time, whereas the equational theory of iterative idempotent semirings is PSPACE-complete, cf. [21].

[*]Partially supported by the project TÁMOP-4.2.1/B-09/1/KONV-2010-0005 "Creating the Center of Excellence at the University of Szeged", supported by the European Union and co-financed by the European Regional Fund, by the TÁMOP-4.2.2/08/1/2008-0008 program of the Hungarian National Development Agency, and by grant no. K 75249 from the National Foundation of Hungary for Scientific Research.

The aim of the present paper is to extend some results from [11] to "iterative multi-linear $K$-$\Sigma$-algebras", which are both $\Sigma$-algebras and $K$-semialgebras for a fixed commutative semiring $K$, satisfying certain natural linearity conditions. The main result shows that when $K$ is proper, the free iterative multi-linear $K$-$\Sigma$-algebras may be described by rational tree series [8]. The result applies to the semiring $\mathbb{N}$ of natural numbers, all finite commutative semirings, all Noetherian commutative semirings and in fact to all commutative semirings $K$ such that every finitely generated subsemiring of $K$ is contained in a Noetherian subsemiring.

The unique fixed point rule has been the basis of several complete axiomatization results in Computer Science. In addition to automata and languages, it has frequently been used in concurrency, see [16, 17, 5], for example. Our freeness result yields a family of such completeness results for rational tree series (regular weighted tree languages) [8, 10].

## 2 Iterative multi-linear algebras

A *semiring* $S = (S, +, \cdot, 0, 1)$ consists of a commutative monoid $(S, +, 0)$ and a monoid $(S, \cdot, 1)$ such that product distributes over all finite sums including the empty sum, so that $x \cdot 0 = 0 = 0 \cdot x$ for all $x \in S$. Examples of semirings include all rings, the semiring $\mathbb{N}$ of natural numbers and the Boolean semiring $\mathbb{B}$ over the set $\{0, 1\}$ whose sum operation is disjunction and whose product operation is conjunction. A *commutative semiring* is a semiring whose product is commutative.

Suppose that $K = (K, +, \cdot, 0, 1)$ is a commutative semiring and $\Sigma$ is a ranked set. A *multi-linear $K$-$\Sigma$-algebra* $B$ is both a $\Sigma$-algebra with operations $\sigma^B : B^n \to B$, for all $\sigma \in \Sigma_n$, $n \geq 0$, and a $K$-semimodule $(B, +, 0)$ with left $K$-action $K \times B \to B$ subject to the usual laws, where $x, y, z \in B$ and $k, k' \in K$:

$$
\begin{align}
(x + y) + z &= x + (y + z) \tag{1} \\
x + y &= y + x \tag{2} \\
x + 0 &= x \tag{3} \\
(k + k')x &= kx + k'x \tag{4} \\
0x &= 0 \tag{5} \\
k(x + y) &= kx + ky \tag{6}
\end{align}
$$

$$k0 = 0 \tag{7}$$
$$(kk')x = k(k'x) \tag{8}$$
$$1x = x \tag{9}$$

Moreover, the following identities hold for all $\sigma \in \Sigma_n$, $i \in [n] = \{1, \ldots, n\}$, $n > 0$ and $k \in K$:

$$\sigma(x_1, \ldots, x_i + y_i, \ldots, x_n) = \sigma(x_1, \ldots, x_i, \ldots, x_n) + \sigma(x_1, \ldots, y_i, \ldots, x_n) \tag{10}$$
$$\sigma(x_1, \ldots, kx_i, \ldots, x_n) = k\sigma(x_1, \ldots, x_n). \tag{11}$$

Morphisms of multi-linear $K$-$\Sigma$-algebras are both $\Sigma$-algebra morphisms and $K$-semimodule morphisms. Below we will often denote the operations $\sigma^B$ of an algebra $B$ by just $\sigma$.

In the sequel, we will be considering systems of fixed point equations with two sorts of variables, namely variables that occur on the left sides of the equations and possibly on the right sides, and variables that only occur on the right sides of the equations. We will call the latter as *parameters*.

The set of all $K$-$\Sigma$-*algebra terms* over the set $X = \{x_1, \ldots, x_n\}$ of variables and the set $A$ of parameters is given by the context-free grammar

$$t ::= x_i \mid a \mid 0 \mid t + t \mid kt \mid \sigma(t, \ldots, t)$$

where $k$ ranges over the semiring $K$, $x_i$ is a variable and $a$ is a parameter. When $\sigma \in \Sigma_0$, we identify the term $\sigma()$ with the letter $\sigma$.

An interpretation of a set $A$ of parameters in a multi-linear $K$-$\Sigma$-algebra $B$ assigns an element of $B$ to each parameter $a$ in $A$, usually also denoted $a$. Given a multi-linear $\Sigma$-algebra $B$ together with an interpretation of the parameters $A$ as elements of $B$, each term $t$ over $\{x_1, \ldots, x_n\}$ in the parameters $A$ induces a *polynomial function* $t^B : B^n \to B$ defined in the usual way.

With respect to the axioms of multi-linear $K$-$\Sigma$-algebras (1) – (11), each term $t$ over the variables in $X$ and parameters $A$ is equivalent to a term of the form $\sum_{i=1}^{m} k_i t_i$, where each $k_i$ is a nonzero element of $K$ and each $t_i$ is a term not containing any occurrence of $+, 0$ and the $K$-action. (When $m = 0$, this term is 0.) Below we will usually identify terms that are equivalent under the axioms of multi-linear $K$-$\Sigma$-algebras.

**Definition 2.1** *Suppose that $X$ is a set of variables and $A$ is an alphabet of parameters. We call a term $t$ over $X$ in the parameters $A$ **proper** (or **guarded**), if it is either $0$, or a letter $a$ in $A$, or a term of the form $\sigma(t_1, \ldots, t_n)$, $\sigma \in \Sigma_n$, $n \geq 0$, where $t_1, \ldots, t_n$ are any terms over $X$ in the parameters $A$, or a term of the form $kt$ or $t_1 + t_2$, where $t, t_1, t_2$ are proper terms and $k \in K$.*

Since for any term $t$, the term $0t$ can be identified with $0$, we could as well allow proper terms of the form $0t$, where $t$ is any term.

When $t$ is a proper term, it is equivalent to a term $\sum_{i=1}^{m} k_i t_i$ where each $k_i$ is a nonzero element of $K$ and each $t_i$ is a term not containing any occurrence of $+, 0$ and the $K$-action which is not a single variable.

**Definition 2.2** *Suppose that $B$ is a multi-linear $K$-$\Sigma$-algebra. We say that $B$ is an **iterative multi-linear $K$-$\Sigma$-algebra** if for each sequence $t_1, \ldots, t_n$ of proper terms over a set $X = \{x_1, \ldots, x_n\}$ of variables and a set $A$ of parameters interpreted as elements of $B$, the finite system of proper fixed point equations*

$$\begin{aligned} x_1 &= t_1 \\ &\vdots \\ x_n &= t_n \end{aligned} \quad (12)$$

*has a unique solution over $B$. Morphisms of iterative multi-linear $K$-$\Sigma$-algebras are just multi-linear $K$-$\Sigma$-algebra morphisms.*

Thus, if $B$ is an iterative multi-linear $K$-$\Sigma$-algebra, then for any system of fixed point equations (12) there is a unique $n$-tuple $(b_1, \ldots, b_n) \in B^n$ with $b_i = t_i^B(b_1, \ldots, b_n)$ for all $i$. When $E$ denotes the system (12), we denote the unique solution $(b_1, \ldots, b_n)$ by $|E|^B$.

We end this section with an example. Suppose that $A$ is a set. A $\Sigma$-*tree* over $A$ is just a $\Sigma$-term generated from the letters in $A$ by the context-free grammar

$$t ::= a \mid \sigma(t, \ldots, t)$$

where $a$ ranges over $A$ and $\sigma \in \Sigma_n$, $n \geq 0$. Let $T_\Sigma(A)$ denote the collection of all $\Sigma$-trees over $A$. Then $T_\Sigma(A)$ is naturally a $\Sigma$-algebra, the free $\Sigma$-algebra over $A$. A $\Sigma$-*tree series* [10] over $A$ with coefficients in a semiring $K$ is a function $s : T_\Sigma(A) \to K$, sometimes denoted as a formal sum $\sum_{t \in T_\Sigma(A)} (s, t) t$, where $(s, t) = s(t) \in K$. Let $K \langle\!\langle T_\Sigma(A) \rangle\!\rangle$ denote the set of all series. We may equip $K \langle\!\langle T_\Sigma(A) \rangle\!\rangle$ by the pointwise sum operation and the pointwise action, so that it becomes a $K$-semimodule with neutral element the constant zero series, denoted 0. For each $\sigma \in \Sigma_n$, $n \geq 0$ and for each $n$-tuple of series $s_1, \ldots, s_n$, let us define $s = \sigma(s_1, \ldots, s_n)$ by

$$(s, t) = \begin{cases} \prod_{i=1}^{n} (s_i, t_i) & \text{if } t = \sigma(t_1, \ldots, t_n) \\ 0 & \text{otherwise.} \end{cases}$$

We identify each $\Sigma$-tree $t$ with the series mapping $t$ to 1 and all other trees to 0. When $K$ is commutative, $K \langle\!\langle T_\Sigma(A) \rangle\!\rangle$ becomes a multi-linear $K$-$\Sigma$-algebra, and in fact an iterative multi-linear $K$-$\Sigma$-algebra, see e.g., [7].

The above construction may be generalized. Call a $\Sigma$-algebra $B$ *finitely decomposable* if for any $\sigma \in \Sigma_n$, each $b \in B$ can be written only a finite number of different ways as $b = \sigma(b_1, \ldots, b_n)$, where $b_1, \ldots, b_n \in B$. Then $K \langle\!\langle B \rangle\!\rangle$, the set of all series $B \to K$ may be equipped with the pointwise sum and $K$-action, and the $\Sigma$-operations

$$(\sigma(s_1, \ldots, s_n), b) = \sum_{b = \sigma(b_1, \ldots, b_n)} \prod_{i=1}^{n} (s_i, b_i)$$

for all $\sigma \in \Sigma_n$, $n \geq 0$. When $K$ is commutative, $K\langle\!\langle B \rangle\!\rangle$ is a multi-linear $K$-semialgebra. However, $K\langle\!\langle B \rangle\!\rangle$ need not be an iterative multi-linear $K$-semialgebra.

Call a $\Sigma$-algebra $B$ *saturated* if there is function $\varphi : B \to \mathbb{N}$ such that for all $\sigma \in \Sigma_n$ and $b_1, \ldots, b_n$, $\varphi(\sigma^B(b_1, \ldots, b_n)) > \max\{\varphi(b_i) : i \in [n]\}$. Clearly, $T_\Sigma(A)$ is saturated by the usual *height* function.

**Proposition 2.3** *For any commutative semiring $K$ and saturated finitely decomposable $\Sigma$-algebra $B$, $K\langle\!\langle B \rangle\!\rangle$ is an iterative multi-linear $K$-$\Sigma$-algebra.*

*Proof outline.* Suppose that $E$ is a finite system of proper polynomial fixed point equations over $X = \{x_1, \ldots, x_n\}$ in the parameters $A$ as in (12). Suppose further that each parameter $a \in A$ is interpreted as an element of $B$. We can write the right side $t_i$ of each equation as $t'_i + t''_i$, where the term $t''_i$ does not contain any variable. Each term $t''_i$ evaluates to a series $r_i$ in $K\langle\!\langle B \rangle\!\rangle$.

Suppose that $(s_1, \ldots, s_n)$ is a solution of $E$ over $K\langle\!\langle B \rangle\!\rangle$. Then for each $b \in B$, each $(s_i, b)$ is the sum of $(r_i, b)$ with an element of $K$ that is completely determined by the $(s_j, b')$ with $\varphi(b') < \varphi(b)$. Thus the solution is unique. Also, since each $(s_i, b)$ depends only on $(r_i, b)$ and the values $(s_j, b')$ with $\varphi(b') < \varphi(b)$, we can define the $(s_i, b)$ by induction on $\varphi(b)$ so that $(s_1, \ldots, s_n)$ becomes a solution. □

When $A$ is a set, let $\mathbf{Rel}_A$ denote the semiring of binary relations on $A$ whose sum operation is union and whose product operation is composition of relations. The constants 0 and 1 are the empty relation and the identity relation on $A$. Since $\mathbf{Rel}_A$ is idempotent, we may view it as a $\mathbb{B}$-$\Sigma$-semialgebra, where $\Sigma$ consists of just the symbol · for multiplication and the constant 1. However, it is not iterative, since fixed point equations may have many solutions.

## 3 Some constructions

Let $K$ denote a commutative semiring. In this section we define descriptions denoting "rational" elements of an iterative $K$-$\Sigma$-algebra. Flat descriptions will correspond to weighted tree automata. Then we provide several (standard) constructions on descriptions and derive from these constructions the fact that the rational elements of an iterative $K$-$\Sigma$-algebra themselves form an iterative $K$-$\Sigma$-algebra.

**Definition 3.1** *Suppose that $D$ is a finite system of polynomial fixed point equations over the variables $X = \{x_1, \ldots, x_n\}$ and the parameters $A$. Moreover, suppose that $v = (v_1, \ldots, v_n) \in K^n$. Then we call the pair $(v, D)$ a **description** over $X_n$ in the parameters $A$. A description $(v, D)$ is called **flat** if the right side of each equation of $D$ is (equivalent to) a finite sum of subterms of the form $k\sigma(x_{i_1}, \ldots, x_{i_m})$ or $ka$, where $k \in K$ and $\sigma \in \Sigma_m$, $m \geq 0$, $i_1, \ldots, i_m \in [n]$ and $a \in A$. When $B$ is an iterative multi-linear $K$-$\Sigma$-algebra together with an interpretation of the parameters in $A$ as elements of $B$, then the **behavior** of $(v, D)$ over $B$ is $|(v, D)|^B = \sum_{i=1}^n v_i |D|_i^B \in B$,* where $|D|_i^B$ denotes the ith component of $|D|^B$, for all $i \in [n]$. Two descriptions are **equivalent over** $B$ if they have the same behavior over $B$ under each interpretation of the parameters. Moreover, two descriptions are **equivalent**, if they are equivalent over all iterative multi-linear $K$-$\Sigma$-algebras.

**Proposition 3.2** *For each description $(v, D)$ there is an equivalent description $(w, E)$ such that the first component of $w$ is 1 and all other components are 0. When $(v, D)$ is flat, then so is $(w, D)$.*

*Proof.* When $D$ is a finite system of fixed point equations over $X = \{x_1, \ldots, x_n\}$ in the parameters $A$ and if $v = (k_1, \ldots, k_n)$, then let the first equation of $E$ be $z = \sum_{i=1}^{n} k_i t_i$, where $t_1, \ldots, t_n$ respectively denote the right sides of the equations for $x_1, \ldots, x_n$ and $z$ is a new variable. The other equations of $E$ are those of $D$. When $(v, D)$ is flat, $(w, E)$ can be turned into an equivalent flat description by using the axioms of $K$-semimodules. □

A flat description $(v, D)$ over $X = \{x_1, \ldots, x_n\}$ in the parameters $A$ may be seen as the *weighted tree automaton* whose set of states is $\{q_1, \ldots, q_n\}$, say, where $q_i$ corresponds to $x_i$, for all $i \in [n]$, which has a transition

$$(q_{i_1}, \ldots, q_{i_m}) \to q_i$$

labeled $\sigma \in \Sigma_m$ of weight $k$ iff $k\sigma(x_{i_1}, \ldots, x_{i_m})$ is a summand of the right side of the $i$th equation of $D$. When $ka$ is a summand of this term, then $q_i$ is $a$-initial with weight $k$. Finally, each state $q_i$ is final with weight $v_i$, the $i$th component of $v$.

Let us interpret each parameter $a \in A$ as the series in $K\langle\!\langle T_\Sigma(A)\rangle\!\rangle$ associated with $a$ that maps $a$ to 1 and all other trees to 0. Then the behavior of $(v, D)$ over $K\langle\!\langle T_\Sigma(A)\rangle\!\rangle$ is just the behavior of the corresponding weighted tree automaton, cf. [8].

In the rest of this section we provide some constructions on descriptions that will be used later in the sequel. Suppose that $(v, D)$ and $(w, E)$ are descriptions over the disjoint sets of variables $X = \{x_1, \ldots, x_m\}$, $Y = \{y_1, \ldots, y_n\}$ in the parameters $A$. Then we define $(v, D) + (w, E)$ to be the description $((v, w), (D, E))$ over $X \cup Y$ in the parameters $A$. Moreover, for each $k \in K$, we define $k(v, D)$ to be the description $(kv, D)$ over $X$.

**Proposition 3.3** *For any iterative multi-linear $K$-$\Sigma$-algebra $B$ with an interpretation of each parameter in $A$ as an element of $B$, we have*

$$|(v, D) + (w, E)|^B = |(v, D)|^B + |(w, E)|^B$$
$$|k(v, D)|^B = k|(v, D)|^B$$

*If $(v, D)$ and $(w, E)$ are flat, then so are $(v, D) + (w, E)$ and $k(v, D)$.*

Suppose now that $\sigma \in \Sigma_n$, and let $(v_1, D_1), \ldots, (v_n, D_n)$ be descriptions over the pairwise disjoint (ordered) sets of variables $X_1, \ldots, X_n$ in the parameters $A$. Without loss of generality we may assume that the first component of each $v_i$ is 1 and the other components of each $v_i$ are 0. Then we let $\sigma(D_1, \ldots, D_n)$ denote the system of fixed point equations whose first component is $z = \sigma(x_{11}, \ldots, x_{n1})$, where each $x_{i1}$ is the first variable of $X_i$ and $z$ is a new variable. The remaining equations are those of the $D_i$, $i = 1, \ldots, n$. Finally, we define $\sigma((v_1, D_1), \ldots, (v_n, D_n))$ to be the description $(v, \sigma(D_1, \ldots, D_n))$ over $\bigcup_{i=1}^n X_i$ in the parameters $A$, where the first component of $v$ (corresponding to the variable $z$) is 1 and its other components are all 0. Note that if each $(v_i, D_i)$ is flat, then so is $\sigma((v_1, D_1), \ldots, (v_n, D_n))$. In particular, when $\sigma \in \Sigma_0$, then the corresponding description is $(v_\sigma, D_\sigma)$, where $D_\sigma$ consists of the single equation $z = \sigma$, and the unique component of $v_\sigma$ is 1. In a similar way, we can associate a description (over the empty set of variables) $(v_0, D_0)$ with 0, and a description $(v_a, D_a)$ with each $a \in A$.

**Proposition 3.4** *In any iterative multi-linear $K$-$\Sigma$-algebra $B$ with an interpretation of the parameters $A$ as elements of $B$ it holds that*

$$\begin{aligned}
|\sigma((v_1, D_1), \ldots, (v_n, D_n))|^B &= \sigma^B(|(v_1, D_1)|^B, \ldots, |(v_n, D_n)|^B) \\
|(v_0, D_0)|^B &= 0 \\
|(v_a, D_a)|^B &= a
\end{aligned}$$

**Definition 3.5** *Suppose that $B$ is an iterative multi-linear $K$-$\Sigma$-algebra with an interpretation of the parameters $A$ as elements of $B$. Then we let $\mathrm{Rat}_B(A)$ denote the set of all $b \in B$ that are the behavior of some description in the parameters $A$ over $B$ with respect to the given interpretation.*

In particular, when $B$ is $K\langle\!\langle T_\Sigma(A) \rangle\!\rangle$, then we denote $\mathrm{Rat}_B(A)$ by $K^{\mathrm{rat}}\langle\!\langle T_\Sigma(A) \rangle\!\rangle$. In the next section, we will characterize $K^{\mathrm{rat}}\langle\!\langle T_\Sigma(A) \rangle\!\rangle$ as the free iterative $K$-$\Sigma$-semialgebra on $A$.

**Proposition 3.6** *Suppose that $B$ is an iterative multi-linear $K$-$\Sigma$-algebra with an interpretation of the parameters $A$ as elements of $B$. Then for each $b \in \mathrm{Rat}_B(A)$ there is a flat description in the parameters $A$ whose behavior over $B$ is $b$.*

*Proof.* Suppose that $b = |(v, D)|$ and write each term on the right side of an equation of $D$ in the form $\sum_{i=1}^m k_i t_i$ where the terms $t_i$ do not contain any occurrence of $+, 0$ or a $K$-action and $k_i \neq 0$ for all $i$ and no $t_i$ is a single variable. If some $t_i$ is of the form $\sigma(s_1, \ldots, s_n)$ where some $s_i$ is not a variable, then we replace $s_i$ with a new variable $z$, say, and introduce the equation $z = s_i$. By continuing this process, we end up with an equivalent flat description, where the initial weights corresponding to the newly introduced variables are all 0.

For example, when the initial description is

$$((k_1, k_2), \ (x_1 = k\sigma(\sigma(a, x_1), x_2) + k'\sigma(x_1, x_2), \ x_2 = \sigma(b, x_1)))$$

the equivalent flat description is

$$((k_1, k_2, 0, 0), E)$$

where $E$ is the following system of equations:

$$\begin{aligned}
x_1 &= k\sigma(x_3, x_2) + k'\sigma(x_1, x_2) \\
x_2 &= \sigma(x_4, x_1) \\
x_3 &= \sigma(x_5, x_1) \\
x_4 &= b \\
x_5 &= a.
\end{aligned}$$

□

**Theorem 3.7** *For every iterative multi-linear $K$-$\Sigma$-algebra $B$ with an interpretation of the parameters $A$ as elements of $B$, $\mathrm{Rat}_B(A)$ is an iterative multi-linear $K$-$\Sigma$-algebra. It is the least iterative multi-linear $K$-$\Sigma$-subalgebra of $B$ containing $A$.*

*Proof.* The fact that $\mathrm{Rat}_B(A)$ is a multi-linear $K$-$\Sigma$-algebra is clear from Propositions 3.3 and 3.4. We prove that it is an iterative multi-linear $K$-$\Sigma$-algebra. For this reason, suppose that $A'$ is a set of parameters with an interpretation as elements of $\mathrm{Rat}_B(A)$ and $(v, D)$ is a description over $X = \{x_1, \ldots, x_n\}$ in the parameters $A'$ whose behavior over $B$ is $b$, say. Without loss of generality we may assume that each parameter $a \in A'$ is in fact in $B$ and is interpreted by itself. For each $a \in A'$ there exists a description $(v_a, D_a)$ over some set $X_a = \{x_1^a, \ldots, x_{n_a}^a\}$ of variables in the parameters $A$ with $|(v_a, D_a)|^B = a$. Let $v_a = (v_1^a, \ldots, v_{n_a}^a)$ and let $D'$ be the system of equations obtained from $D$ by replacing each occurrence of a parameter $a \in A'$ by $\sum_{i=1}^{n_a} v_i^a t_i^a$, where for each $i$, $t_i^a$ denotes the right side of the $i$th equation of $D_a$. Finally, let $a_1, \ldots, a_k$ be all the parameters that occur in $D$, and consider the description

$$((v, 0, \ldots, 0), (D', D_{a_1}, \ldots, D_{a_k})).$$

The behavior of this description (in the parameters $A$) over $B$ is clearly $b$.

The fact that $\mathrm{Rat}_B(A)$ is the least multi-linear $K$-$\Sigma$-subalgebra of $B$ containing the interpretations of the parameters in $A$ is clear by definition. □

## 4 Simulations

In this section, following and [6, 13], we introduce simulations between descriptions, and show that descriptions that can be connected by a sequence of simulations are equivalent. Let $K$ denote a commutative semiring.

Suppose that $D$ and $E$ are finite systems of proper polynomial fixed point equations over $X = \{x_1, \ldots, x_m\}$ and $Y = \{y_1, \ldots, y_n\}$ in the parameters $A$. Let $x_i = s_i$ and $y_j = t_j$, $i \in [m], j \in [n]$ denote the equations of $D$ and $E$, respectively. When $M \in K^{m \times n}$, we define $DM$ to be the system of equations obtained by replacing each occurrence of each variable $x_i$ on the right side of an equation of $D$ by $M_{i1}y_1 + \ldots + M_{in}y_n$:

$$x_1 = s_1(M_{11}y_1 + \ldots + M_{1n}y_n, \ldots, M_{m1}y_1 + \ldots + M_{mn}y_n)$$
$$\vdots$$
$$x_m = s_m(M_{11}y_1 + \ldots + M_{1n}y_n, \ldots, M_{m1}y_1 + \ldots + M_{mn}y_n)$$

Moreover, we let $ME$ denote the system of equations:

$$x_1 = M_{11}t_1 + \ldots + M_{1n}t_n$$
$$\vdots$$
$$x_m = M_{m1}t_1 + \ldots + M_{mn}t_n$$

**Definition 4.1** *Suppose that $(v, D)$ and $(w, E)$ are descriptions over $X = \{x_1, \ldots, x_m\}$ and $Y = \{y_1, \ldots, y_n\}$ in the set of parameters $A$. A **simulation** $(v, D) \to (w, E)$ is given by a matrix $M \in K^{m \times n}$ such that $vM = w$ and the right side of each equation of $DM$ is equivalent to the right side of the corresponding equation of $ME$ with respect to the axioms of multi-linear $K$-$\Sigma$-algebras.*

Word automata connected by a simulation are called *conjugate* in [2, 3].

**Theorem 4.2** *Suppose that $(v, D)$ and $(w, E)$ are descriptions over $X = \{x_1, \ldots, x_m\}$ and $Y = \{y_1, \ldots, y_n\}$ in the set of parameters $A$ as above. Moreover, suppose that $M$ is a simulation $(v, D) \to (w, E)$. Then $(v, D)$ and $(w, E)$ are equivalent.*

*Proof.* Suppose that $B$ is an iterative multi-linear $K$-$\Sigma$-algebra with an interpretation of each parameter in $A$ as an element of $B$, and suppose that $(c_1, \ldots, c_n)$ is the unique solution of $E$ over $B$. Define $b_1, \ldots, b_m$ by $b_i = M_{i1}c_1 + \ldots + M_{in}c_n$ for each $i \in [m]$. Then

$$\begin{aligned} s_i^B(b_1, \ldots, b_n) &= s_i^B(M_{11}c_1 + \ldots + M_{1n}c_n, \ldots, M_{m1}c_1 + \ldots + M_{mn}c_n) \\ &= M_{i1}t_1^B(c_1, \ldots, c_n) + \ldots + M_{in}t_n^B(c_1, \ldots, c_n) \\ &= M_{i1}c_1 + \ldots + M_{in}c_n \\ &= b_i, \end{aligned}$$

for all $i \in [m]$. Thus $(b_1, \ldots, b_m) \in B^m$ is a solution $D$ over $B$, showing that $|D|^B = M|E|^B$. Using this, $|(v, D)|^B = v|D|^B = vM|E|^B = w|E|^B$. □

**Definition 4.3** *We say that a commutative semiring $K$ is **proper** if for all flat $(v, D)$ and $(w, E)$ over $X$ and $Y$ in the parameters $A$, if $(v, D)$ and $(w, E)$ are equivalent over $K\langle\!\langle T_\Sigma(A)\rangle\!\rangle$ when each parameter $a$ is interpreted as the corresponding tree series, then there is a sequence of simulations connecting them, i.e., there is a sequence $(v_1, D_1), \ldots, (v_k, D_k)$ of descriptions in the parameters $A$ and simulations $M_i : (v_i, D_i) \to (v_{i+1}, D_{i+1})$ or $(v_{i+1}, D_{i+1}) \to (v_i, D_i)$, $i = 1, \ldots, k-1$ such that $(v_1, D_1) = (v, D)$ and $(v_k, D_k) = (w, E)$.*

In [13], it is shown that all commutative rings are proper. More generally, let us call a (commutative) semiring *Noetherian*, if for every finitely generated $K$-semimodule $M$, every subsemimodule of $M$ is finitely generated. It is shown in [13] that if $K$ is Noetherian, or every finitely generated subsemiring of $K$ is contained in a Noetherian subsemiring, then $K$ is proper. In particular, every locally finite[1] (commutative) semiring and thus every bounded distributive lattice is proper. There exist proper semirings that are not Noetherian, for example the semiring $\mathbb{N}$, cf. [3, 13]. An example of a semiring that is not proper is the "tropical semiring", cf. [12].

**Corollary 4.4** *Suppose that $K$ is proper. Then for any* flat $(v, D)$ *and* $(w, E)$ *over $X$ and $Y$ in the parameters $A$, if $(v, D)$ and $(w, E)$ are equivalent over $K\langle\!\langle T_\Sigma(A) \rangle\!\rangle$ when each parameter $a$ is interpreted as the tree series corresponding to $a$, then $(v, D)$ and $(w, E)$ are equivalent.*

## 5  Freeness

In this section we prove our main result:

**Theorem 5.1** *Suppose that $K$ is proper commutative semiring. Then for each set $A$, $K^{\mathrm{rat}}\langle\!\langle T_\Sigma(A) \rangle\!\rangle$ is the free iterative multi-linear $K$-$\Sigma$-semialgebra on $A$.*

*Proof.* We already know (cf. Theorem 3.7) that $K^{\mathrm{rat}}\langle\!\langle T_\Sigma(A) \rangle\!\rangle$ is an iterative multi-linear $K$-$\Sigma$-algebra. Let $\eta$ denote the map (interpretation) that assigns to each $a \in A$ the corresponding series that maps $a$ to 1 and all other trees in $T_\Sigma(A)$ to 0. Each series in $K^{\mathrm{rat}}\langle\!\langle T_\Sigma(A) \rangle\!\rangle$ is the behavior of a (flat) description in the parameters $A$ under this interpretation. Since morphisms of multi-linear $K$-$\Sigma$-algebras preserve unique solutions of proper polynomial fixed point equations, it follows that for any iterative multi-linear $K$-$\Sigma$-algebra $B$ and function $h : A \to B$, there is at most one morphism $h^\sharp : K^{\mathrm{rat}}\langle\!\langle T_\Sigma(A) \rangle\!\rangle \to B$ extending $h$. Indeed, when $s = |(v, D)|^{K^{\mathrm{rat}}\langle\!\langle T_\Sigma(A) \rangle\!\rangle}$ for a description $(v, D)$ in the parameters $A$ under the interpretation $\eta$, we are forced to define $h^\sharp(s)$ as the behavior $|(v, D)|^B$ of $(v, D)$ over $B$ with respect to the interpretation that maps each parameter $a$ to $h(a)$.

In the rest of the proof, we show that an extension exists. So let $h : A \to B$. When $s \in K^{\mathrm{rat}}\langle\!\langle T_\Sigma(A) \rangle\!\rangle$, there is a *flat* description $(v, D)$ in the parameters $A$ whose behavior over $K^{\mathrm{rat}}\langle\!\langle T_\Sigma(A) \rangle\!\rangle$ with respect to the interpretation $\eta$ is the series $s$. We define $h^\sharp(s) := |(v, D)|^B$, the behavior of the same description over $B$ with respect to the interpretation $h$.

Since $K$ is proper, if $(w, E)$ is another flat description whose behavior over $K^{\mathrm{rat}}\langle\!\langle T_\Sigma(A) \rangle\!\rangle$ with respect to $\eta$ is $s$, then by Corollary 4.4, $|(v, D)|^B = |(w, E)|^B$. Thus, the definition of $h^\sharp(s)$ does not depend on the choice of the flat description $(v, D)$ whose behavior over $K^{\mathrm{rat}}\langle\!\langle T_\Sigma(A) \rangle\!\rangle$ with respect to $\eta$ is $s$.

---

[1] A semiring is locally finite if its finitely generated subsemirings are finite.

By these facts, and the constructions given above, it follows now that $h^\sharp$ is a morphism. We only need to show that $h^\sharp$ preserves the $\Sigma$-operations. To this end, let $\sigma \in \Sigma_n$, where $n \geq 0$, and let $s_1, \ldots, s_n \in K^{\text{rat}}\langle\!\langle T_\Sigma(A)\rangle\!\rangle$, $s = \sigma(s_1, \ldots, s_n)$. For each $i$, let $(v_i, D_i)$ be a flat description whose behavior over $K^{\text{rat}}\langle\!\langle T_\Sigma(A)\rangle\!\rangle$ with respect to $\eta$ is $s_i$. Let $(v, D) = \sigma((v_1, D_1), \ldots, (v_n, D_n))$, so that $(v, D)$ is also flat. Then the behavior of $(v, D)$ over $K^{\text{rat}}\langle\!\langle T_\Sigma(A)\rangle\!\rangle$ is $s$. By definition, $h^\sharp(s_i) = |(v_i, D_i)|^B$, for all $i \in [n]$, and $h^\sharp(s) = |(v, D)|^B$. But by Proposition 3.4, $|(v, D)|^B = \sigma^B(|(v_1, D_1)|^B, \ldots, |(v_n, D_n)|^B)$, so that $h^\sharp(s) = \sigma^B(h^\sharp(s_1), \ldots, h^\sharp(s_n))$. □

Theorem 5.1 applies to $\mathbb{N}$ and to all commutative semirings $K$ such that every finitely generated subsemiring of $K$ is included in a Noetherian subsemiring. In particular, Theorem 5.1 applies to all commutative rings and to all locally finite commutative semirings including $\mathbb{B}$. When $K$ is $\mathbb{B}$, Theorem 5.1 implies the axiomatization of regular $\Sigma$-tree languages by the unique fixed point rule [14], and when additionally $\Sigma$ consists of unary letters and $A$ is a singleton set, then Theorem 5.1 implies Salomaa's axiomatization [20] of regular word languages by the unique fixed point rule.

With some obvious modifications, Theorem 5.1 may be extended to *commutative iterative multi-linear $K$-$\Sigma$-algebras* satisfying the identity

$$\sigma(x_1, \ldots, x_n) = \sigma(x_{1\pi}, \ldots, x_{n\pi})$$

for all $\sigma \in \Sigma_n$ and for all permutations $\pi : [n] \to [n]$. One needs to replace $\Sigma$-trees by "unordered" $\Sigma$-trees.

# 6 Conclusion

We may consider each set $\Sigma$ as a ranked set all of whose letters have rank 1. Consequently, we may identify a word $u \in \Sigma^*$ with a tree $t \in T_\Sigma(\{a\})$, where $a$ is a fixed parameter. Under this identification, regular languages in $\Sigma^*$ correspond to regular tree languages in $T_\Sigma(A)$, and vice versa. Salomaa's axiomatization of regular word languages may be formulated as the assertion that for any (finite) set $\Sigma$, $\mathbb{B}^{\text{rat}}\langle\!\langle T_\Sigma(\{a\})\rangle\!\rangle$ is freely generated by the letter $a$ in the class of all iterative $\mathbb{B}$-$\Sigma$-algebras. In our main contribution, Theorem 5.1, we have extended Salomaa's characterization to $K$-$\Sigma$-algebras $K^{\text{rat}}\langle\!\langle T_\Sigma(A)\rangle\!\rangle$, where $K$ is a proper commutative semiring, $\Sigma$ is any ranked set and $A$ is any set of free generators. Proper commutative semirings include $\mathbb{N}$, all locally finite commutative semirings and all commutative rings.

# References

[1] J. Adámek, S. Milius and J. Velebil, Iterative algebras at work. *Math. Structures Comput. Sci.*, 16(2006), 1085–1131.

[2] M.-P. Béal, S. Lombardy, J. Sakarovitch, On the equivalence of Z-automata. In: Proc. ICALP 2005, LNCS 3580, Springer, 2005, 397–409.

[3] M.-P. Béal, S. Lombardy and J. Sakarovitch, Conjugacy and equivalence of weighted automata and functional transducers. In: *Proc. CSR 2006*, LNCS 3967, Springer, 2006, 58–69.

[4] D. Benson and I. Guessarian, Iterative and recursive matrix theories. *J. Algebra*, 86(1984), 302–314.

[5] J. A. Bergstra and J. W. Klop, A complete inference system for regular processes with silent moves. In: *Proceedings, Logic Colloquium 1886*, (F. R. Drake and J. K. Truss, Eds.), North-Holland, Amsterdam, 1988, 21–81.

[6] S. L. Bloom and Z. Ésik, *Iteration Theories*, Springer, 1993.

[7] S. L. Bloom and Z. Ésik, An extension theorem with an application to formal tree series, *J. of Automata, Languages and Combinatorics,* 8(2003), 145–185.

[8] M. Droste, W. Kuich, H. Vogler (eds.), *Handbook of Weighted Automata*, Springer, 2009.

[9] C. C. Elgot, Monadic computation and iterative algebraic theories. In: *Logic Colloquium '73, (Bristol, 1973), Studies in Logic and the Foundations of Mathematics, Vol. 80*, North-Holland, Amsterdam, 1975, 175–230.

[10] Z. Ésik and W. Kuich, Formal tree series, *J. Automata, Languages and Combinatorics*, 8(2003), 219–285.

[11] Z. Ésik and W. Kuich, Free iterative and iteration $K$-semialgebras, *Alg. Universalis*, to appear.

[12] Z. Ésik and A. Maletti, Simulation vs. equivalence. In: *Proc. 6th Int. Conf. Foundations of Computer Science, FCS 2010*, (Las Vegas, Nevada), CSREA Press, 119–122.

[13] Z. Ésik and A. Maletti, Simulations of tree automata. In: Proc. *15th Int. Conf. Implementation and Application of Automata*, Winnipeg, Canada, 2010, LNCS, Springer, to appear.

[14] T. Ito and S. Ando, A complete axiom system of super-regular expressions. In: *Information processing 74* (IFIP Congress, Stockholm), North-Holland, Amsterdam, 1974, 661–665.

[15] D. Krob, Rational expressions over a ring. In: *Topics in Invariant Theory* (Paris, 1989/1990), Lecture Notes in Math., 1478, Springer, Berlin, 1991, 215–243,

[16] R. Milner, A complete inference system for a class of regular behaviours. *J. Comput. System Sci.*, 28(1984), 439–466.

[17] R. Milner, A complete axiomatisation for observational congruence of finite-state behaviours. *Inform. and Comput.*, 81(1989), 227–247.

[18] M. Morisaki and K. Sakai, A complete axiom system for rational sets with multiplicity. *Theoret. Comput. Sci.*, 11(1980), 79–92.

[19] E. Nelson, Iterative algebras. *Theoret. Comput. Sci.*, 25(1983), 67–94.

[20] A. Salomaa, Two complete axiom systems for the algebra of regular events. *J. Assoc. Comput. Mach.*, 13(1966), 158–169.

[21] L. J. Stockmeyer and A. R. Meyer, Word problems requiring exponential time. In: proc. *STOC '73*, ACM, 1973, 1–9.

[22] J. Tiuryn, Unique fixed points vs. least fixed points. *Theoret. Comput. Sci.*, 12(1980), no. 3, 229–254.

[23] W. Wechler, Iterative nondeterministic algebras. *Studia Automat.*, 13(1989), 35–44.